\documentclass[smallextended]{svjour3}       
\usepackage{amssymb}
\usepackage{bm}
\usepackage{latexsym}
\usepackage{amsfonts,amssymb}
\usepackage{graphicx,epsfig}
\usepackage{psfrag}
\usepackage{amsmath,amssymb}

\def\s {\sigma}

\def\be {\begin{equation}}
\def\ee {\end{equation}}
\def\br{\begin{eqnarray}}
\def\er{\end{eqnarray}}
\def\bc {\begin{center}}
\def\ec {\end{center}}
\def\bfg {\begin{figure}}
\def\efg {\end{figure}}
\def\bi {\begin{itemize}}
\def\ei {\end{itemize}}
\def\benu{\begin{enumerate}}
\def\eenu{\end{enumerate}}
\newcommand{\bdm}{\begin{displaymath}}
\newcommand{\edm}{\end{displaymath}}

\def\nn {\nonumber}

\def\la {\label}
\def\le {\left}
\def\ri {\right}

\def\f {\frac}
\def\sq {\sqrt}
\def\s {\sigma}

\def\hs {\hspace}

\def\laq{\hbox{~}\raise 0.4ex\hbox{$<$}\kern -0.8em\lower 0.62ex\hbox{$\sim$}\hbox{~}}
\def\gaq{\hbox{~}\raise 0.4ex\hbox{$>$}\kern -0.7em\lower 0.62ex\hbox{$\sim$}\hbox{~}}

\newsavebox{\blambox}\savebox{\blambox}[0.6em]{$\lambda\!\!\!$\raisebox{0.5em}
{$\neg$}}
\newsavebox{\bFox}\savebox{\bFox}[0.6em]{$F\!\!\!\!$\raisebox{0.5em}
{$\neg$}}
\newsavebox{\bxibox}\savebox{\bxibox}[0.6em]{$\xi\!\!\!$\raisebox{.5em}
{$\neg$}}

\DeclareMathOperator{\sech}{sech}

\begin{document}


\title{5D non-symmetric gravity and geodesic confinement} 

\author{Suman Ghosh and  S. Shankaranarayanan}
\institute{School of Physics, Indian Institute of Science Education and 
Research, Thiruvananthapuram 695016, India}

\maketitle
\begin{abstract}
This work focuses on an unexplored aspect of non-symmetric geometry where {\it only} the off-diagonal metric components along the extra dimension, in a 5-dimensional spacetime, are non-symmetric. We show that the energy densities of the stationary non-symmetric models are similar to that of brane models thereby mimicking the thick-brane scenario. We find that the massive test particles are confined near the location of the brane for both growing and decaying warp factors. This feature is unique to the non-symmetric nature of our model. We have also studied the dynamical models where standard 4D FLRW brane is embedded. Our analysis shows that the non-symmetric terms deconfine energy density at the early universe while automatically confine at late times.
\keywords{Non-symmetric metric, extra dimension, geodesic confinement, thick branes.}
\end{abstract}

\vspace*{2.0cm}


\section{Introduction}

General relativity (GR) has survived numerous experimental tests \cite{Will-lrr} and achieved the status of the accepted theory of gravitation. Though observations, relating to galactic rotation curves \cite{Feng,Hooper} and supernova data \cite{Perlmutter} can only be explained within the framework of GR with exotic matter fields. These lead to inception of many new ideas in cosmology as attempts to justify the observational facts. 
In the literature, broadly, there have been two approaches: (a) Assume GR is correct and include exotic matter fields in the right hand side of Einstein equation (like dark matter or dark energy) and (b) Include the possibility that GR is valid only upto a certain length scale. At largest and smallest length scales,  theory of gravity is modified. Such models can potentially avoid introduction of exotic matter fields and can plausibly explain observations.
Until now, both these approaches have been partially successful to explain observations. However, any modification to GR has to be such that it does not contradict with GR in the regime where it is consistent with observations.

Most of the modifications to GR has focused on introducing higher order curvature terms which dominate at larger length scale or smaller length scale. However very few attempts have been made in the literature to look at the simplest of plausible extensions to the assumption of GR-- non-symmetricity of the affine connection.
It is important to understand that GR assumes that spacetime manifold is metrically connected.  
GR further assumes that the metric tensor is symmetric which eventually leads to symmetric affine connections. Physically this means that the corresponding spacetime is torsion free. However, geometric nature of gravity does not require that. 
There have been attempts to construct, more fundamental, affine gravity models in the literature \cite{Weyl,Schrod,Papa}.

In reference \cite{Schrod}, Schr\"odinger described the field equations of non-symmetric purely affine gravity theory. Moffat has extensively studied the, so called non-symmetric gravitational theory (NGT) for over four decades \cite{M1}-\cite{M7}. One of the key results of NGT models, unlike GR or higher derivative gravity models, the black-hole solutions are singularity free \cite{M1,M2,M4,M5,burko-ori-1995,Poplawski}.  Attempts have been made to constraint Moffat's NGT through observations using the results of recent free-fall Galilean test of the WEP \cite{Will}. However this class of models has some difficulties related to ghost poles \cite{Damour-1993} and non-uniqueness \cite{J-P-2006}. 
Recently, attempts were made to show that NGT can be used as an alternative to dark matter and dark energy \cite{M-JCAP-05,M-2004,P-V-2006}.
Vacuum properties of nonsymmetric gravity in de Sitter space are studied in \cite{J-P-2007}.
Further in \cite{Singh-2010}, potential signatures of noncommutative geometry within the muon decay spectrum near the event horizon of a microscopic Schwarzschild black hole (``where the antisymmetric part of the metric tensor manifests the hypothesized noncommutative geometric structure'').

Models with extra dimensions ($D>4$) have also been studied in detail \cite{KK} and gained substantial attention, in the last few decades, mostly motivated from string theory \cite{string}. In the last decade or so, warped braneworld models are extensively analysed, given its potential to solve some longstanding puzzles in theoretical physics \cite{bw-2010,ADD,RS}. Thick brane solutions (3-brane with, finite width, embedded in a 5D bulk) of Einstein equations in warped spacetimes, which are natural generalisation to original infinitely thin brane scenario of RS, appear in various multidimensional field theories coupled to gravity \cite{thick-brane}. Note that brane thickness is an essential ingredient for extending the idea of brane world to multidimensional spacetimes. 
One of the key assumptions in the brane models is that the $5^{th}$ dimension has a discrete symmetry which distinguishes them from the other 4 dimensions indirectly, explaining the non-observability of the extra dimension. While this is a crucial feature, it is not clear how to include this discrete symmetry from fundamental principles.

In this work, we show that the continuous symmetry i.e. non-symmetric metric for the higher dimension mimics the feature of thick brane scenario. We analytically construct stationary and dynamic solutions of field equations in simple five dimensional non-symmetric bulk spacetimes, where the contracted torsion tensor vanishes, and analyse the corresponding energy densities and timelike geodesic motion. We assume that the (off-diagonal) metric components along the extra dimension is anti-symmetric while the metric components of the 4D part is symmetric.

We find that the geometry itself confines the timelike geodesics near the location of the brane for both growing and decaying warp factors. This feature is different from the symmetric bulk models where confinement happens only in presence of growing warp factor \cite{SG-SK-HN}. The dynamic solutions has the standard 4D FLRW universe embedded in the 5D bulk. 

This paper is organised as follows. In section II, we briefly review and introduce the field equations in non-symmetric gravity. In sections III and IV, we analyse the stationary and dynamic solutions respectively. Section V has a brief discussion over possibilities of finding dynamic warped braneworld solutions. Then we conclude with a summary of the results found in Section VI. Signature convention used is $(-1,1,1,1,1)$. Capital Latin indices, Greek indices and small Latin indices are used to denote the coordinates of 5D spacetime, 4D spacetime and 3D space respectively. Ordinary and covariant differentiations are denoted by a coma and a semi-colon, respectively. An overdot ($\dot{}$) means differentiation w.r.t. the affine parameter $\lambda$.

\section{Non-symmetric geometry}

The gravitational field equations in non-symmetric spacetime, in $(D+1)$ dimensions, are given by \cite{Schrod} 
\br
(\sqrt{-g}g^{[AB]})_{,B} &=& 0, \la{eq:1}\\
g_{{}_{AB,C}} - \tilde\Gamma^D_{{}_{AC}}g_{{}_{DB}} - \tilde\Gamma^D_{{}_{CB}}g_{{}_{AD}}  &=& 0 \la{eq:2}\\
G_{{}_{AB}} \equiv R_{{}_{AB}} - \f{1}{2} g_{{}_{AB}} R &=& 8 \pi T_{{}_{AB}}
\la{eq:3}
\er
where $g_{{}_{AB}}$ and $T_{{}_{AB}}$ are the metric  and matter tensors respectively. `$[*]$' means anti-symmetrisation (`$(*)$' will mean symmetrisation). $\tilde\Gamma^D_{{}_{AC}}$'s are the modified affine connections which are {\it recursively} related to the affine connections $\Gamma^A_{{}_{BC}}$ as
\be
\tilde{\Gamma}^A_{{}_{BC}} = \Gamma^A_{{}_{BC}} - \f{2}{D} \delta^A_B ~ {\Gamma}_C \hs{1cm}\mbox{where}\hs{1cm} {\Gamma}_C = \f{1}{2}({\Gamma}^D_{{}_{CD}} - {\Gamma}^D_{{}_{DC}}) \la{eq:modgamma} 
\ee
which implies 
\be
\tilde\Gamma_C = \f{1}{2}(\tilde\Gamma^D_{{}_{CD}} - \tilde\Gamma^D_{{}_{DC}}) = 0 .\la{eq:constraint1}
\ee 
Note that $\Gamma^C_{{}_{[AB]}}$ and $\tilde\Gamma^C_{{}_{[AB]}}$ are both (torsion) tensors, however the connections are not.
$R$ and $R_{{}_{AB}}$ are the Ricci scalar and the first Ricci tensor, respectively, which are defined as:
\be
R = g^{{}_{AB}}R_{{}_{AB}} \hs{1cm}\mbox{where}\hs{1cm} R_{{}_{AB}} = R^C_{{}_{ACB}}
\ee
and the Riemann tensor is given by
\be
R^D_{{}_{ACB}} = \tilde\Gamma^D_{{}_{AB,C}} - \tilde\Gamma^D_{{}_{AC,B}} + \tilde\Gamma^K_{{}_{AB}} \tilde\Gamma^D_{{}_{KC}}  - \tilde\Gamma^K_{{}_{AC}} \tilde\Gamma^D_{{}_{KB}}.
\ee
Note that, since the spacetime is non-symmetric, the ordering of indices in the definition of the tensors is important \cite{Schrod}.

The second Ricci tensor is defined as
\be
Q_{{}_{AB}} = \tilde\Gamma^C_{{}_{CB,A}} - \tilde\Gamma^C_{{}_{CA,B}} \la{eq:2ndricci}
\ee
which is anti-symmetric and has the form of a curl. 
The covariant derivative of a vector $S_{{}_{A}}$ is defined as
\be
S_{{}_{A;B}} = S_{{}_{A,B}} - \Gamma^C_{{}_{A,B}}S_{{}_{C}}.
\ee

Note that Eq. (\ref{eq:constraint1}) is identical to Eq. (\ref{eq:1}). 
Since Eq. (\ref{eq:modgamma}) is recursive, it is difficult to solve analytically. Hence, we assume the following 
\be
\Gamma_C = \f{1}{2}(\Gamma^D_{{}_{CD}} - \Gamma^D_{{}_{DC}}) = 0, \la{eq:constraint2}
\ee
which essentially implies 
\be
\tilde\Gamma^A_{{}_{BC}} = \Gamma^A_{{}_{BC}}. \la{eq:assump}
\ee
This further implies that the second Ricci tensor, Eq. (\ref{eq:2ndricci}), identically vanishes and the corresponding geodesics are also constrained by Eq. (\ref{eq:assump}).
We show that one can construct rich class of interesting solutions, specifically with only one space dimension being non-symmetric, following the steps below:
\begin{enumerate}
\item Assume a simple 5D metric ansatz whose 4D part is symmetric and the off-diagonal components corresponding to the extra dimension is non-symmetric.
\item Derive the affine connections using Eq. (\ref{eq:2}).
\item Then determine the metric functions solving the equations (\ref{eq:1}) or (\ref{eq:constraint1}) or (\ref{eq:constraint2}).
\item Analyse the resulting matter energy density $G_{00}$ assuming that these models are solutions (numerically) of field equations with appropriate matter energy tensor added to the right hand side of Eq.~(\ref{eq:3}).
\item Solutions of the corresponding geodesic equations then reveals how motion of test particles is modified by the non-symmetric nature of the space time. Investigation of whether Weak Energy Condition, $G_{{}_{AB}}u^Au^B>0$, is satisfied or not, tell us to what extent these models are compatible with general relativistic framework.
\end{enumerate}
In the next Section, we discuss stationary spacetimes.

\section{Stationary bulk spacetime}

Let us assume the following simple ansatz for the stationary 5D metric:
\begin{center}
\be
g_{{}_{AB}}  = \le( \begin{array}{lllll} 
-f(\sigma)  & 0  & 0 & 0 & d(\sigma) \\
0 & f(\sigma) & 0 & 0 & d(\sigma) \\
0  & 0 & f(\sigma) & 0 & d(\sigma)  \\
0 & 0 & 0 & f(\sigma) & d(\sigma)  \\
-d(\sigma) & -d(\sigma) & -d(\sigma) & -d(\sigma) & r^2    \\
\end{array} \ri) \la{eq:metric1}
\ee
\end{center}
where the non-zero metric components are functions of the extra spatial dimension $\sigma$. Here $r$ is a constant and in all the figures, presented later, it's value is taken to be unity.
Note that the line element corresponding to metric (\ref{eq:metric1}),
\be
ds^2 = f(\sigma) [-d\eta^2 + d\vec x^2] + r^2 d\sigma^2, \la{eq:lineelement}
\ee
does not capture the true nature of the bulk which is stationary.

Solving Eq.~({\ref{eq:2}}) with our ansatz, the components of $\Gamma^C_{{}_{AB}}$ can be obtained as functions of $f(\sigma)$ and $d(\sigma)$ (see Appendix). Then Eq. (\ref{eq:constraint2}) imply
\be
3\f{d^2}{f^2}f_{,\s} + r^2\le(\f{f_{,\s}}{f} + \f{d_{,\s}}{d}\ri) = 0 . \la{eq:constraint3}
\ee
The above equation has two unknown metric functions, hence, one has to choose one and determine the other. Note that the results found does not depend on the sign of $d(\s)$ since one can always exchange the diagonally opposite anti-symmetric metric components.

We will study two different scenarios -- A) $f(\s) = \cosh^2(\s)$ and B) $f(\s) = \sech^2(\s)$. 

\subsection{$f(\s) = \cosh^2(\s)$}

\begin{figure}
\begin{center}
\includegraphics[scale=.42]{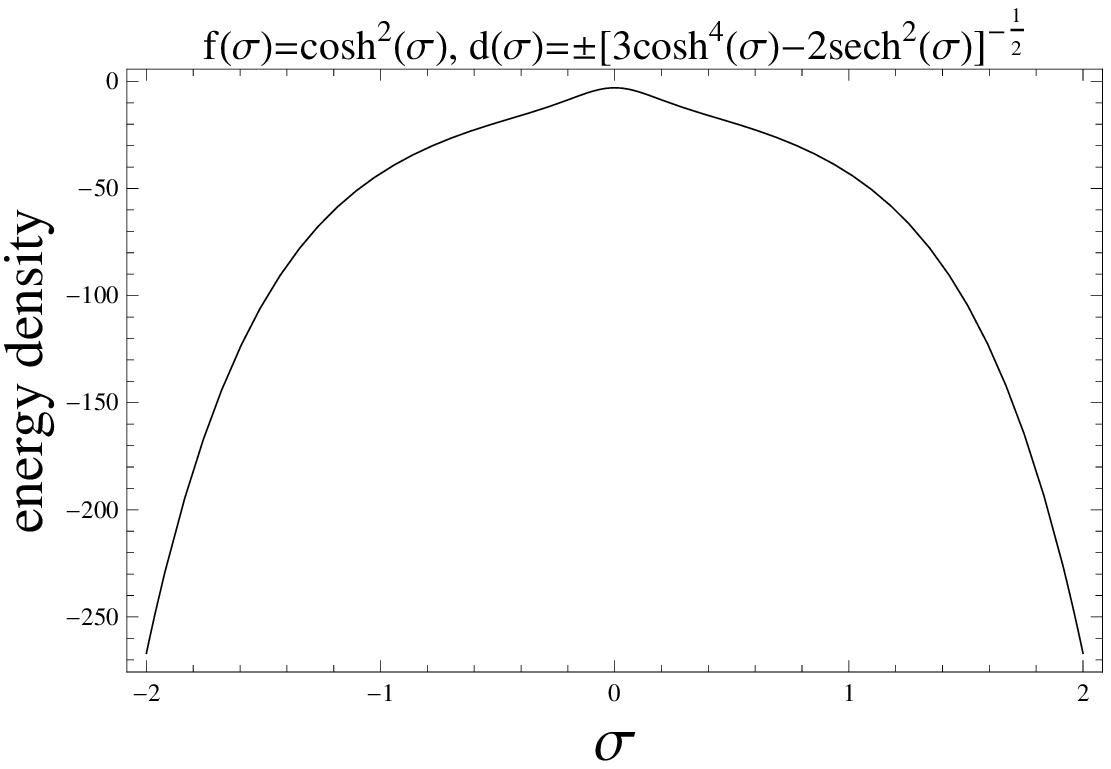}\hs{.5cm}
\includegraphics[scale=.38]{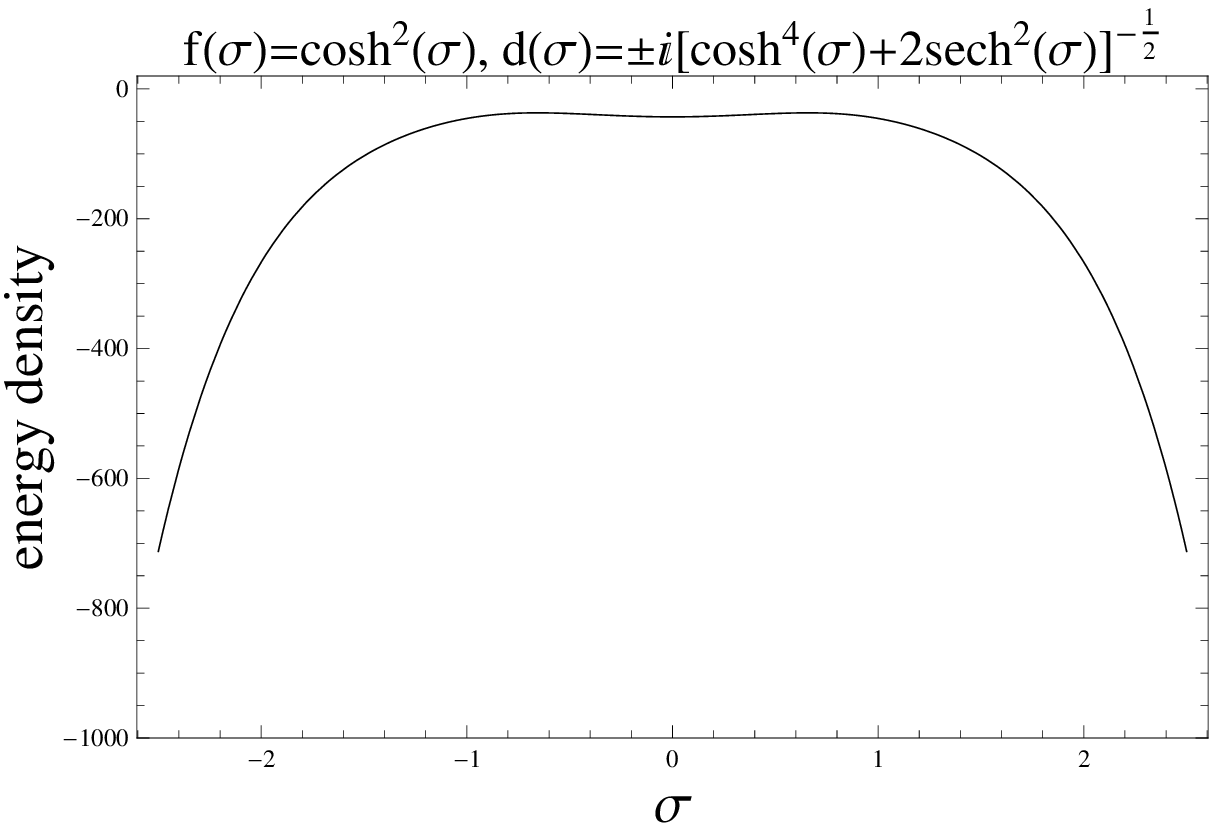} \hs{.5cm}
\includegraphics[scale=.55]{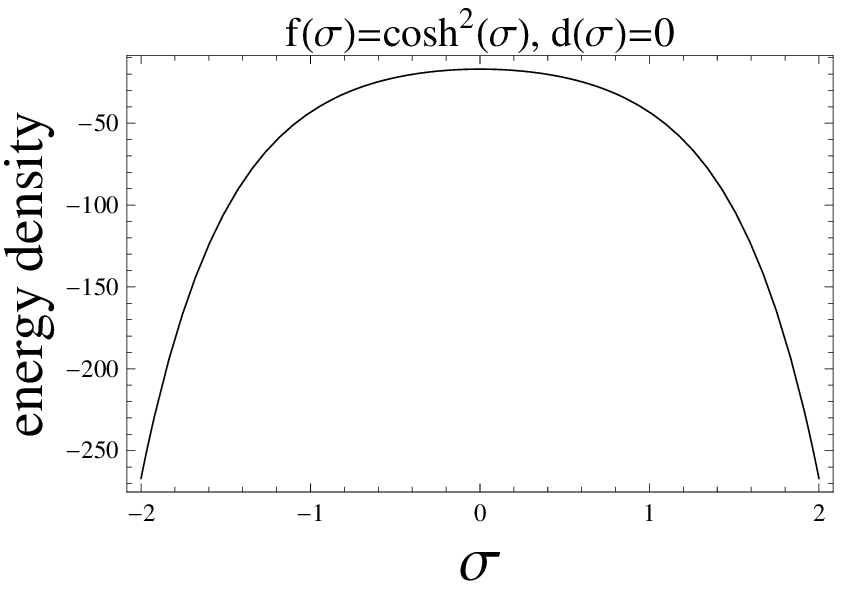} 
\caption{Plots of energy densities vs $\s$ for growing warp factor with $d(\s)$ real, imaginary and zero respectively.}
\label{fig:case1_1}
\end{center}
\end{figure}

Putting $f(\s) = \cosh^2(\s)$ in Eq. (\ref{eq:constraint3}) one gets:
\be
d(\sigma) = \pm \f{r}{\sq{C_1r^2 \cosh^4(\s) - 2\sech^2(\s)}}
\ee
where $C_1$ is the integration constant. Note that $d(\s)$ can be real or purely imaginary, for all $\s$, depending on whether $C_1\ge2/r^2$ or $C_1<0$ respectively. For $0<C_1<2/r^2$, $d(\s)$ is singular at $|\s|<\cosh^{-1}(2/C_1r^2)^{(1/6)}$. Fig. \ref{fig:case1_1} depicts energy density profiles in two such cases where $d(\s)$ is either real ($r=1$, $C_1=3$) or imaginary ($r=1$, $C_1=-1$). 
Energy densities in both the situations behave almost similarly i.e. reaches maximum at $\s=0$. In this sense, these models mimic a thick brane located at $\s=0$. This behavior is not drastically different from the symmetric case because with increasing $|\s|$, $d(\s)$ becomes negligibly small compared to $f(\s)$. 
Note that this case mimics the thick brane scenario, with a {\it growing} warp factor, where the brane is located at $\s=0$.

\subsection{$f(\s) = \sech^2(\s)$}

\begin{figure}[!htb]
\begin{center}
\includegraphics[scale=.44]{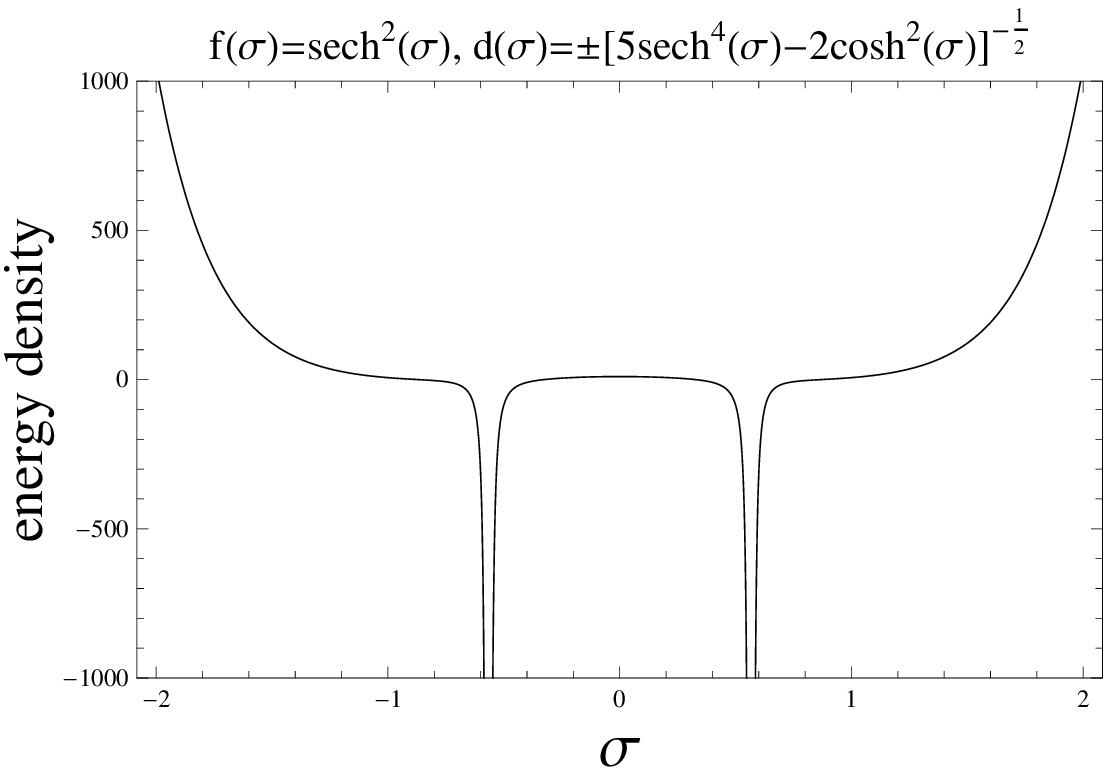} \hs{.5cm}
\includegraphics[scale=.38]{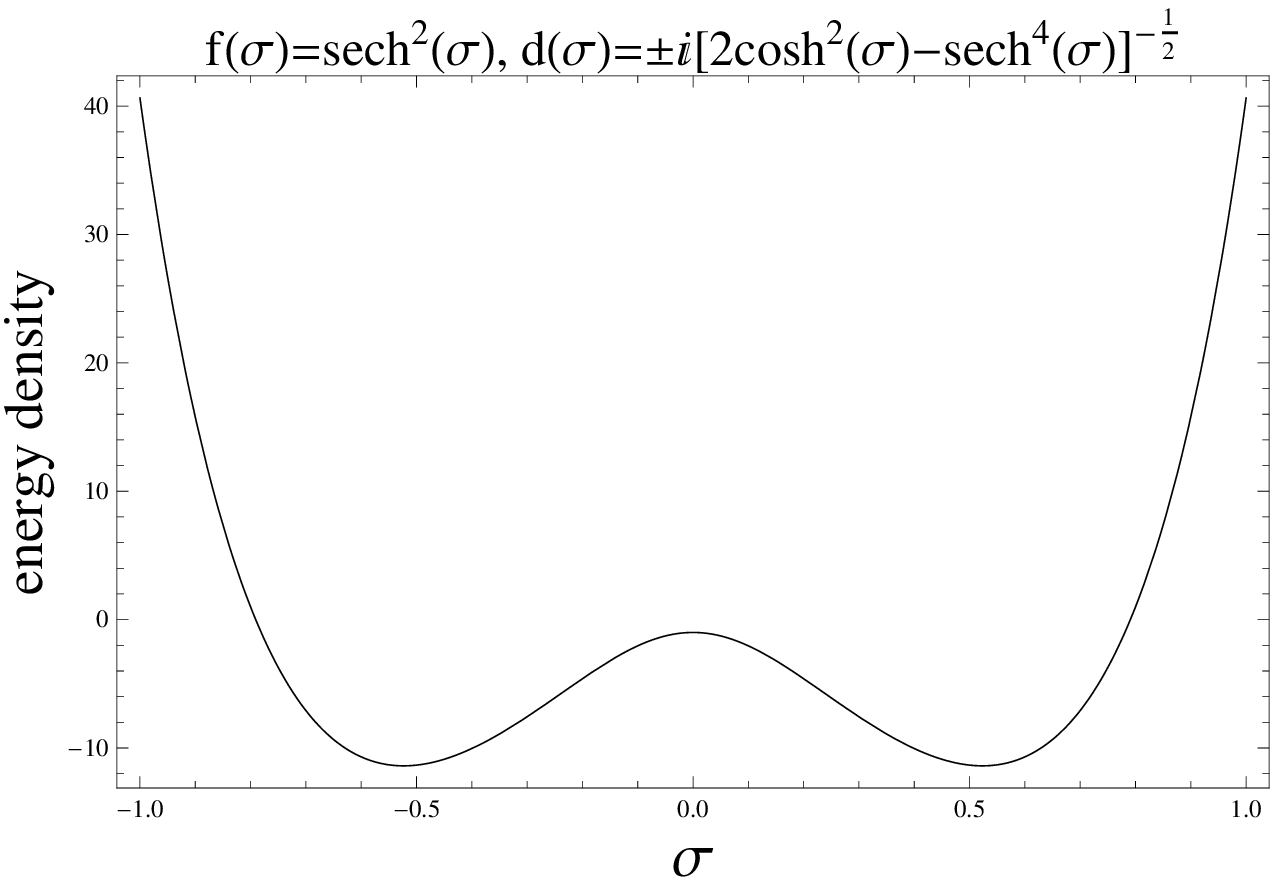} \hs{.5cm}
\includegraphics[scale=.5]{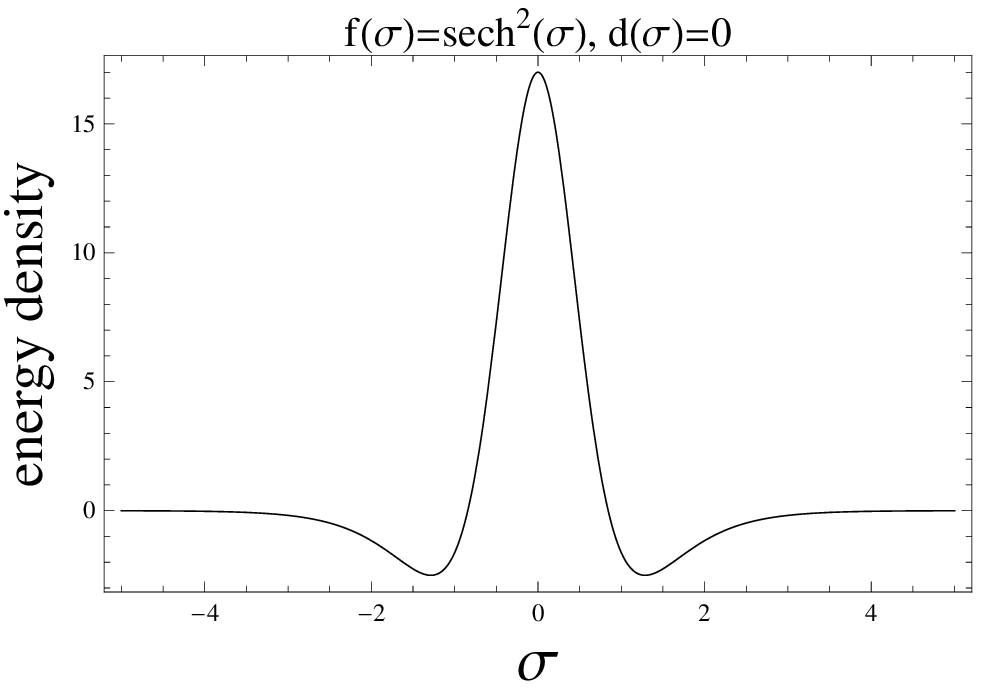} 
\caption{Plots of energy densities vs $\s$ for growing warp factor with $d(\s)$ real, imaginary and zero respectively.}
\label{fig:case2_1}
\end{center}
\end{figure}

With $f(\s) = \sech^2(\s)$, Eq. (\ref{eq:constraint3}) implies:
\be
d(\sigma) = \pm \f{r}{\sq{C_2r^2 \sech^4(\s) - 2\cosh^2(\s)}}
\ee
where $C_2$ is an integration constant. Note that $d(\s)$ is purely imaginary, for all $\s$, when $C_1\leq 2/r^2$. For $C_1>2/r^2$, $d(\s)$ is real below $|\s|=\cosh^{-1}(2/C_1r^2)^{(1/6)}$ where it becomes singular.
Fig. \ref{fig:case2_1} depicts energy density profiles in two such cases where $d(\s)$ has singularities  ($r=1$, $C_2=5$) and purely imaginary ($r=1$, $C_2=-1$). 
It is interesting to note that non-symmetric components does drastically change the energy density profile compared to the symmetric scenario, unlike the previous model where the effect of $d(\s)$ is negligible for large $\s$. This case mimics the thick brane scenario, with a {\it decaying} warp factor, where the brane is located at $\s=0$. Note that when $d(\s)$ is purely imaginary, energy density has a local maxima at $\s=0$ and the location of the brane is less stable compared to the other situations.

In the next section, we look at the corresponding geodesic equations to understand the nature of the spacetime discussed in this section.

\subsection{Geodesics}

The geodesic equations, satisfying the  time-like constraint
\be
f(-\dot\eta^2 + \sum_i \dot x_i^2) + r^2 \dot\s^2 = -1, \la{eq:timelike}
\ee
derived for general ansatz can be written in the following simplified form:
\br
\ddot\eta + \f{f_{,\s}}{f} \dot\eta\dot\s - \f{1}{2}\f{F_{,\s}}{F}(\dot\eta + \sum_i\dot x_i)\dot\s &=& 0, \la{eq:geo1} \\
\ddot x_i + \f{f_{,\s}}{f} \dot x_i\dot\s + \f{1}{2}\f{F_{,\s}}{F}(\dot\eta + \sum_i\dot x_i)\dot\s &=& 0, \la{eq:geo2} \\
\ddot\s + \f{f_{,\s}}{f} \f{(1+\dot\s^2)}{2r^2} - \f{f}{2r^2}\f{F_{,\s}}{F}(\dot\eta + \sum_i\dot x_i)^2 &=& 0, \la{eq:geo3}
\er
where $F=(2d^2/f+r^2)$. Eqs. (\ref{eq:geo1})-(\ref{eq:geo3}) imply that non-symmetric components do effect the geodesic motion although they do not appear in the line element (\ref{eq:lineelement}). Note that as $\eta$ and $x_i$ are the cyclic coordinates, one expects $\dot\eta \propto 1/f$ and $\dot x_i\propto 1/f$. However such solutions does not exist except in special cases when the last terms in the left hand side of the geodesic equations vanish. 

This could occur for three cases-- 
\begin{enumerate}
\item $f\propto d^2$: this condition along with Eq. (\ref{eq:constraint3}) would imply $f,d=constant$.  
\item $\dot\eta =\dot x_i = 0$: implying there is no flow of time which would be unphysical. 
\item $\dot\eta = -\sum_i\dot x_i$: this corresponds to set of geodesics in symmetric spacetime.  In general, putting $F=constant$ scenario refers to the geodesics in a symmetric 5D bulk, that are studied in detail in \cite{SG-SK-HN}. 
\end{enumerate}

Note that in our case behaviour of geodesics can not be understood from the so called geodesic potential \cite{Wald} derivable from the line element as such. As already pointed earlier, effects of non-symmetric nature of the spacetime does not show up in geodesic potential. Here we are interested in the role played by the non-symmetric components of the metric. The equations (\ref{eq:geo1}-\ref{eq:geo3}) are highly coupled and hence difficult to solve analytically. Hence we solve them numerically and plot the results. 

Note that the qualitative aspects of the geodesics do not change with the variation of initial conditions.

\subsubsection{$f(\s) = \cosh^2(\s)$}

\begin{figure}[!htb]
\begin{center}
\includegraphics[scale=.43]{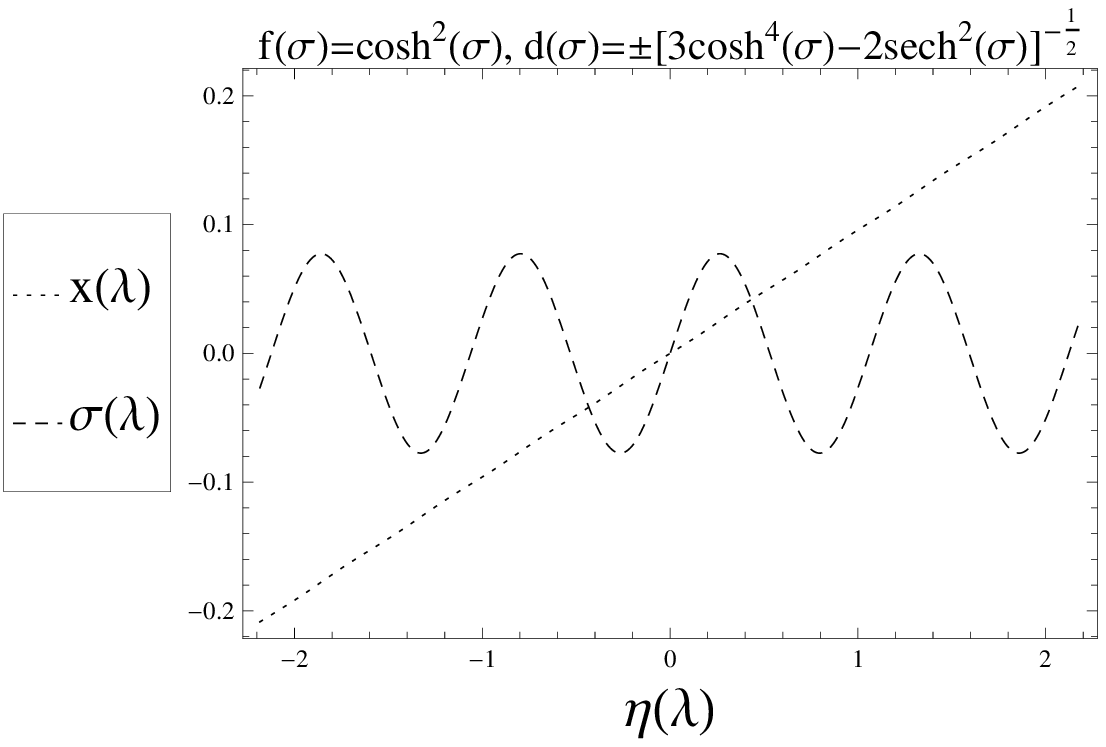} 
\includegraphics[scale=.43]{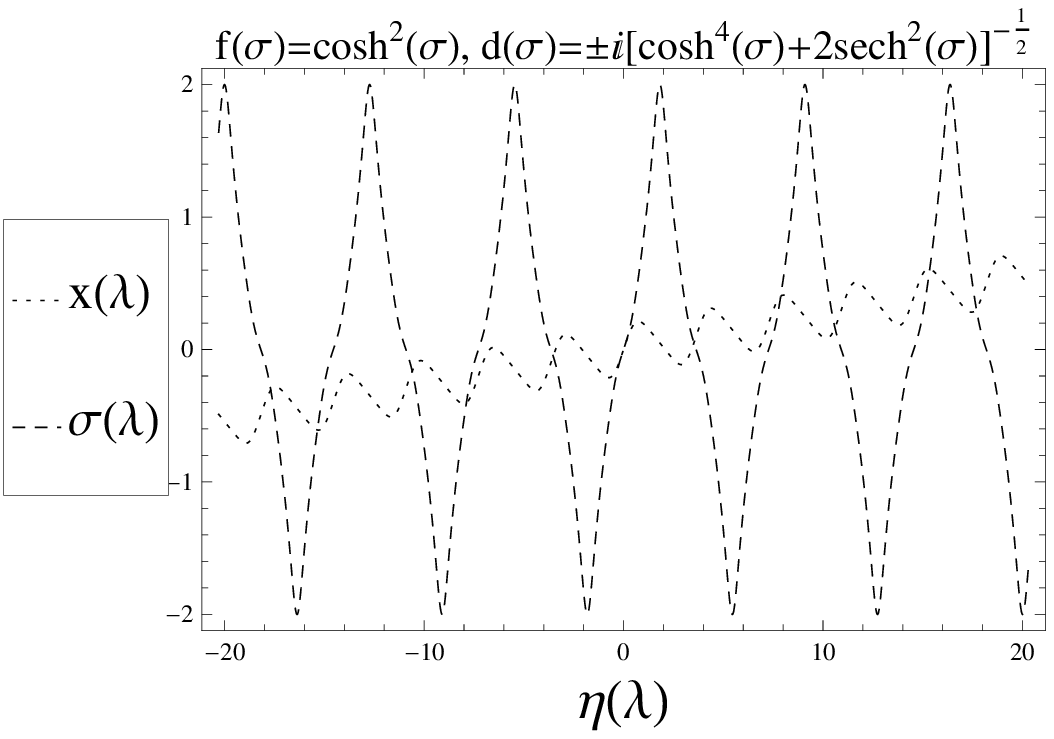} 
\includegraphics[scale=.43]{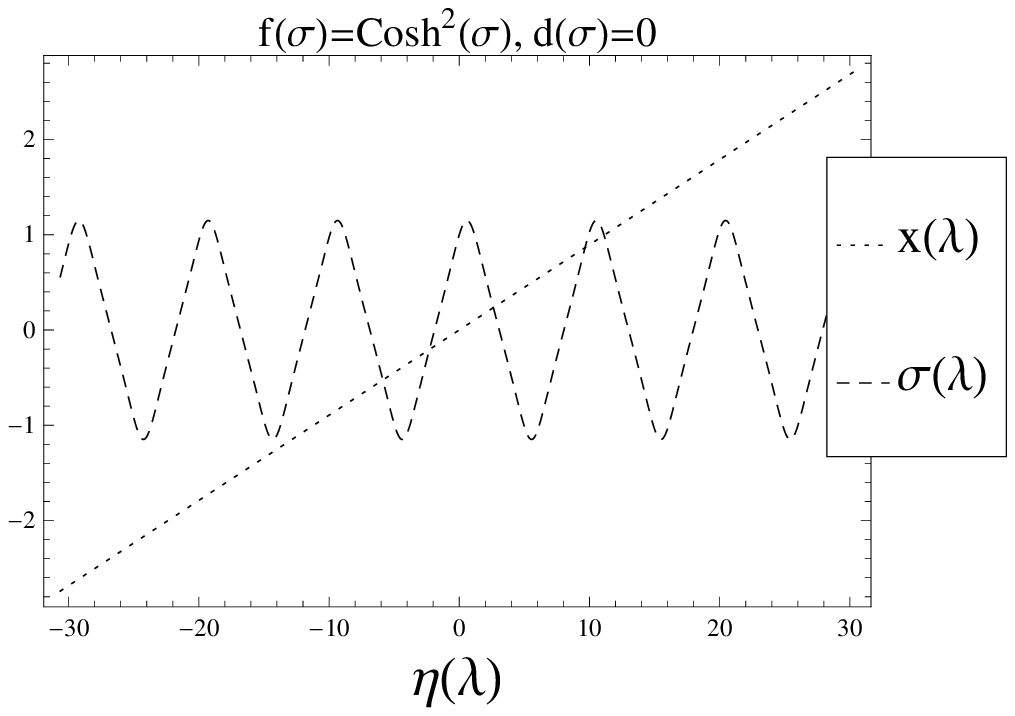} 
\caption{Plot of geodesics $x_i(\lambda)$ and $\sigma(\lambda)$ vs $\eta(\lambda)$ for growing warp factor with $d(\s)$ real, imaginary and zero respectively. Initial conditions are $x_i(0)=\s(0)=0, \eta(0)=0$ and $\dot x_i(0)=\dot\s(0)=1, \dot\eta(0)=\sq{5}$.}
\label{fig:case1_2}
\end{center}
\end{figure}

The geodesics in this growing warp factor scenario are shown in Fig. \ref{fig:case1_2}. In all the cases geodesics look similar and essentially their motion along the extra dimension are confined to $\s=0$. Effect of the non-symmetric nature do not modify the $\s$-component. However when $b(\s)$ is imaginary $x_i$-components also become oscillatory.

\begin{figure}[!h]
\begin{center}
\includegraphics[scale=.5]{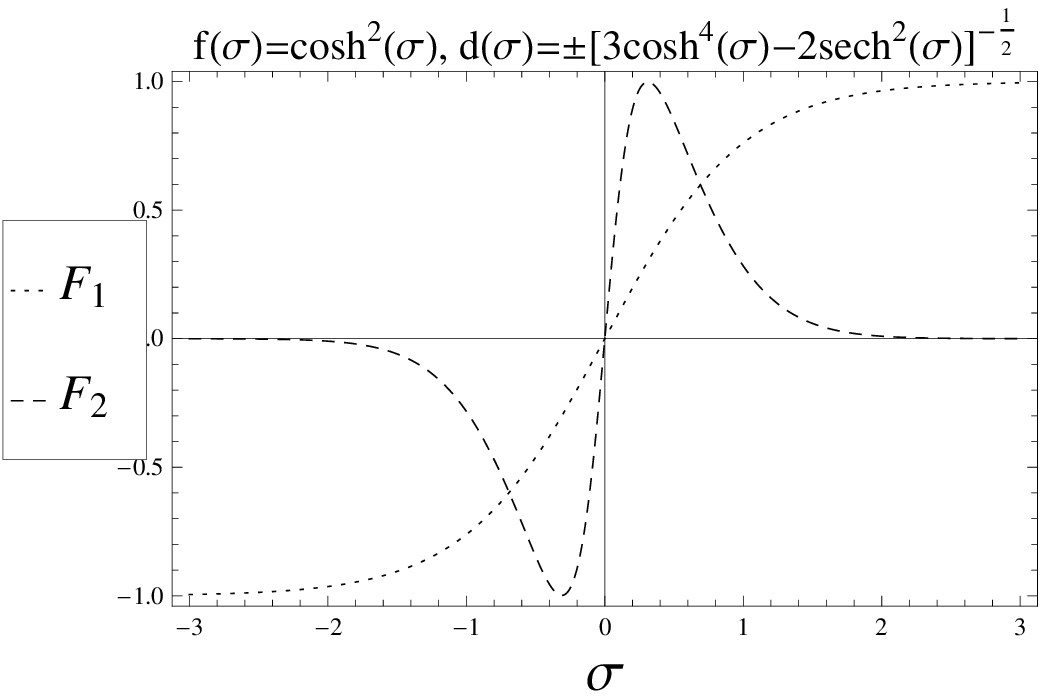} 
\includegraphics[scale=.5]{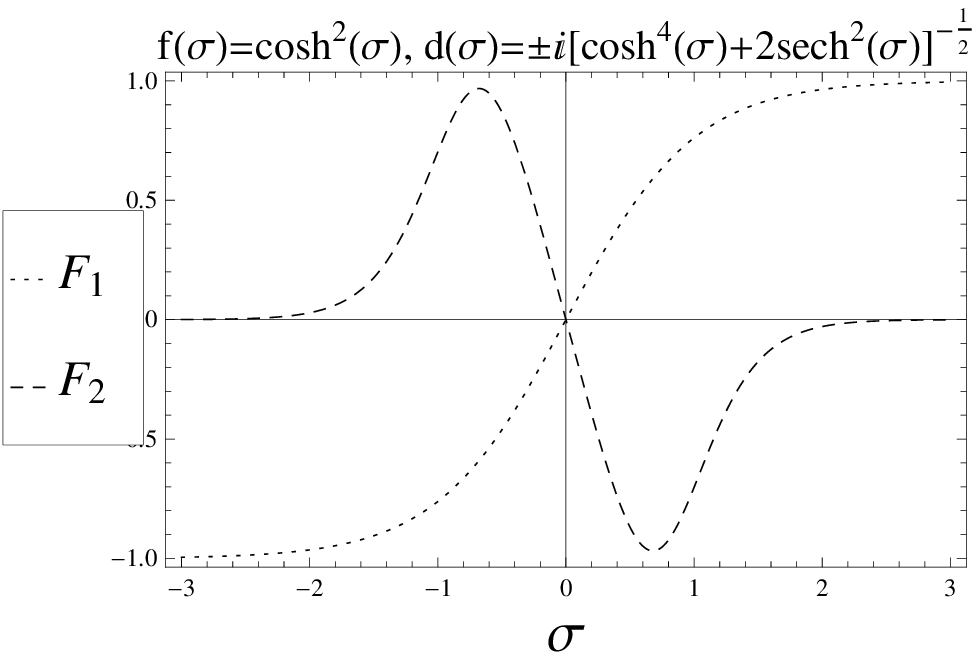} 
\caption{Comparing the coefficients appearing in the geodesic equation where $F_1 = f_{,\s}/2f$ and $F_2 = -ff_{,\s}/2F$.}
\label{fig:case1_F}
\end{center}
\end{figure}

The oscillatory behavior of $\s(\lambda)$ with $d(\s)=0$, can be analytically understood as follows. With growing warp factor Eq. (\ref{eq:geo3}), to the leading order in $\s$ can be written as
\be
\ddot\s + \s \times C_0 = 0 \la{eq:shm}
\ee
where $C_0$ is a positive quantity.
Eq. (\ref{eq:shm}) clearly resembles equation of simple harmonic motion.
However this equation is non-trivial for $d(\s)\neq0$. Numerical analysis shows that the new coefficient appearing in the geodesic equations also contributes in a similar manner in this case.  From  Fig. \ref{fig:case1_F}, the oscillation of $\s(\lambda)$ for real $d(\s)$ seems obvious. Whereas for imaginary $d(\s)$, the oscillatory behaviour of $\s(\lambda)$ stems through $x_i(\lambda)$ as in Eq. (\ref{eq:geo2}), the non-symmetric contribution (or $F_2$ in Fig. \ref{fig:case1_F}), comes with a negative sign. It is worth mentioning here that these reasonings also establish the reliability of the numerical results presented.

Note that the model with real $d(\s)$ satisfy the Weak Energy Condition  whereas for imaginary $d(\s)$, $G_{{}_{AB}}u^Au^B$ become a complex quantity.
\subsubsection{$f(\s) = \sech^2(\s)$}

\begin{figure}[!htb]
\begin{center}
\includegraphics[scale=.39]{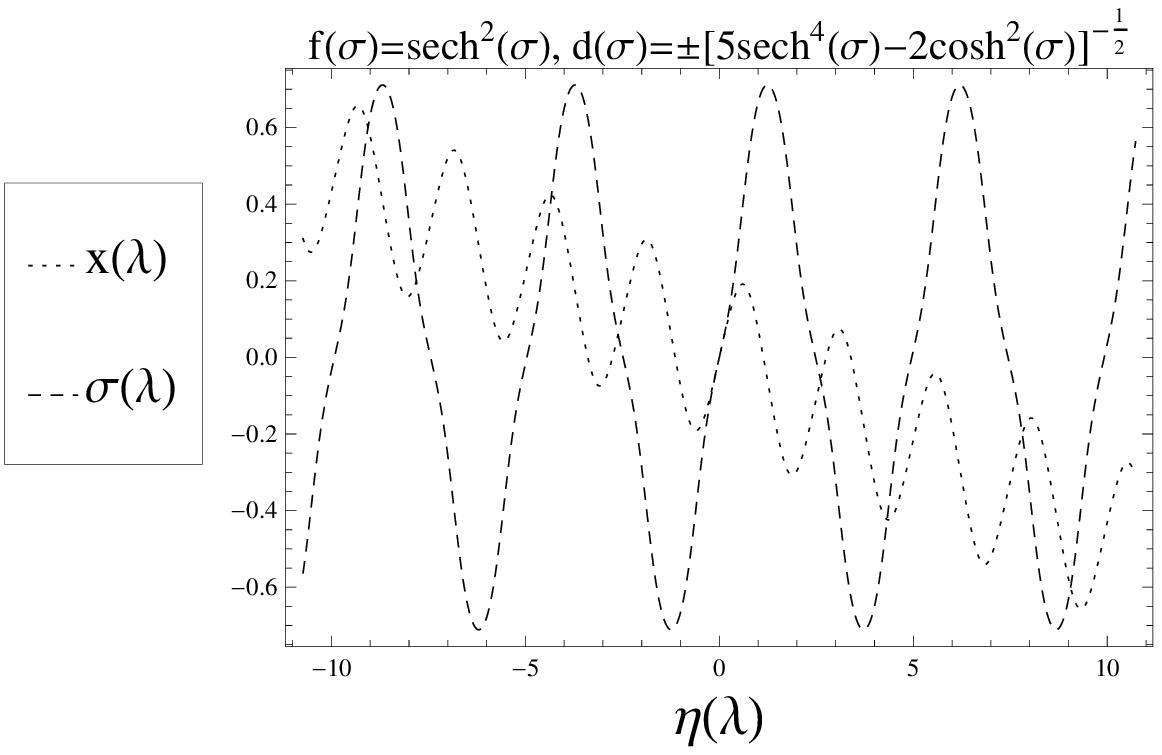} 
\includegraphics[scale=.39]{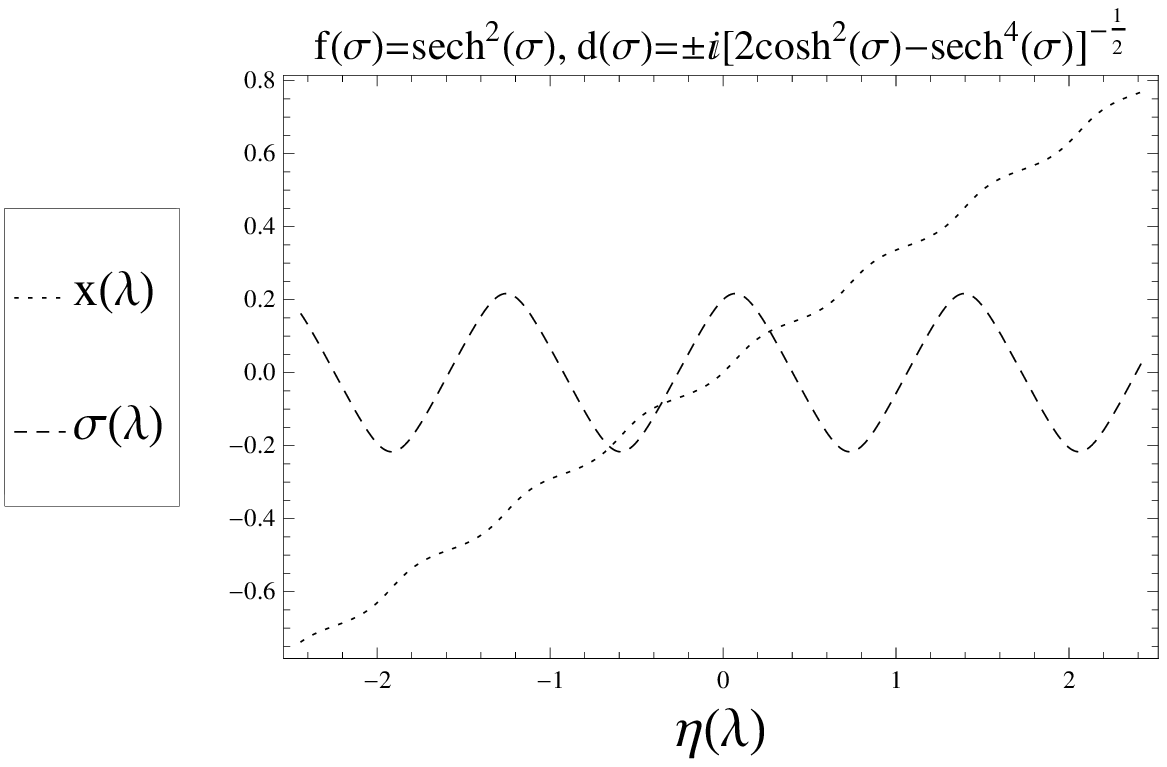} 
\includegraphics[scale=.39]{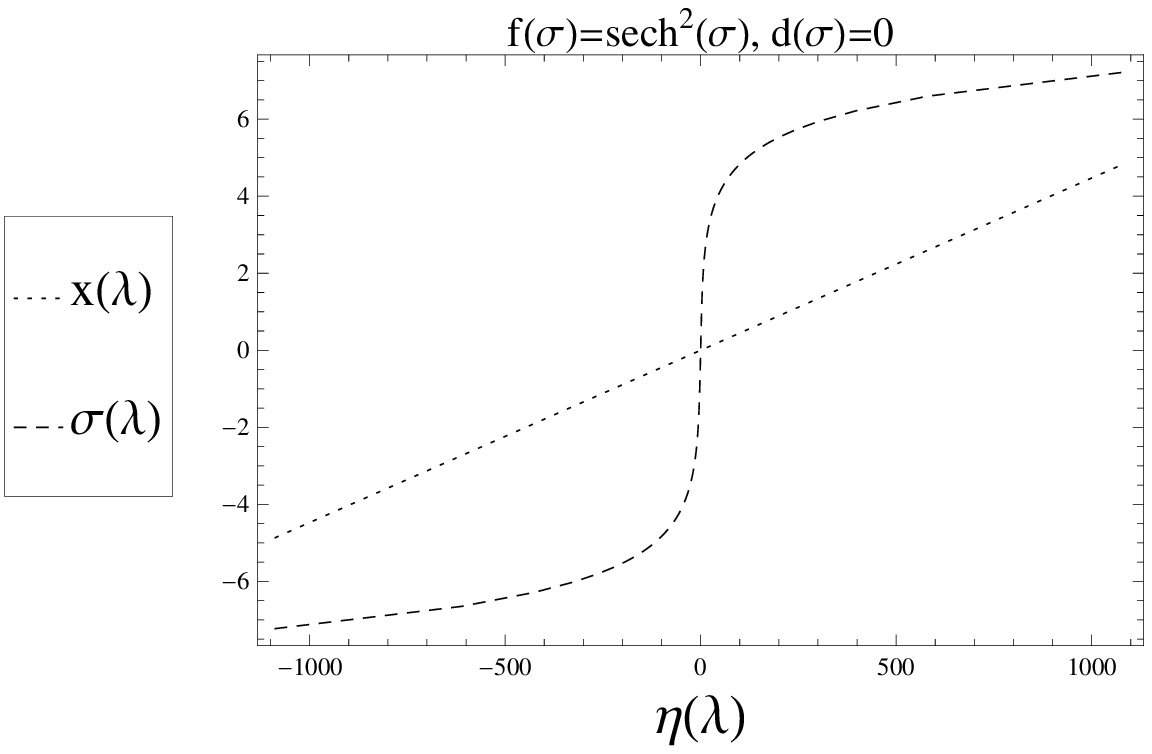} 
\caption{Plots of geodesics $x_i(\lambda)$ and $\sigma(\lambda)$ vs $\eta(\lambda)$ for decaying warp factor with $d(\s)$ real, imaginary and zero respectively. Initial conditions are $x_i(0)=\s(0)=0, \eta(0)=0$ and $\dot x_i(0)=\dot\s(0)=1, \dot\eta(0)=\sq{5}$.}
\label{fig:case2_2}
\end{center}
\end{figure}

\begin{figure}[!h]
\begin{center}
\includegraphics[scale=.5]{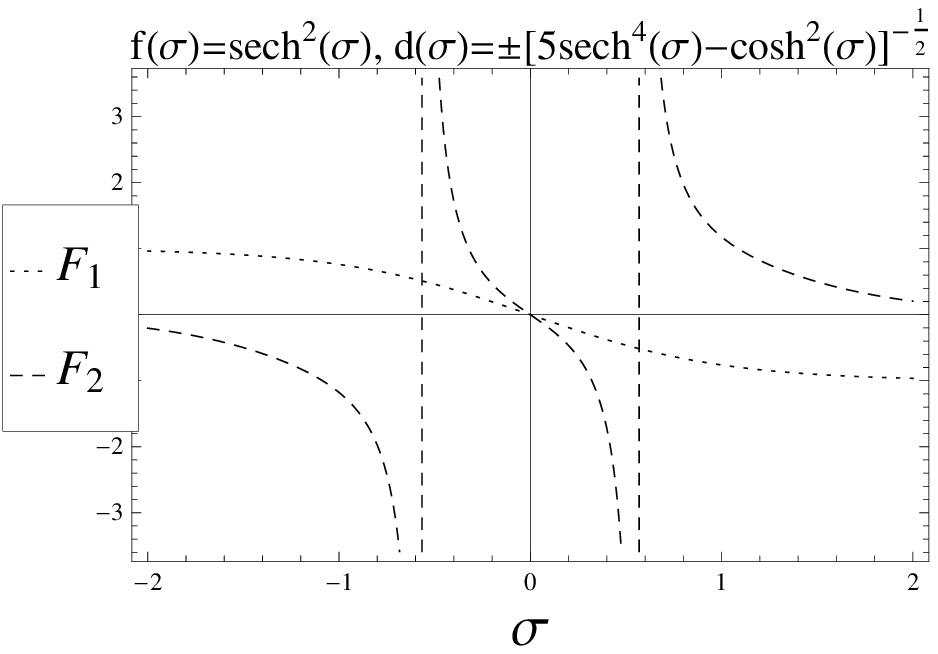} 
\includegraphics[scale=.5]{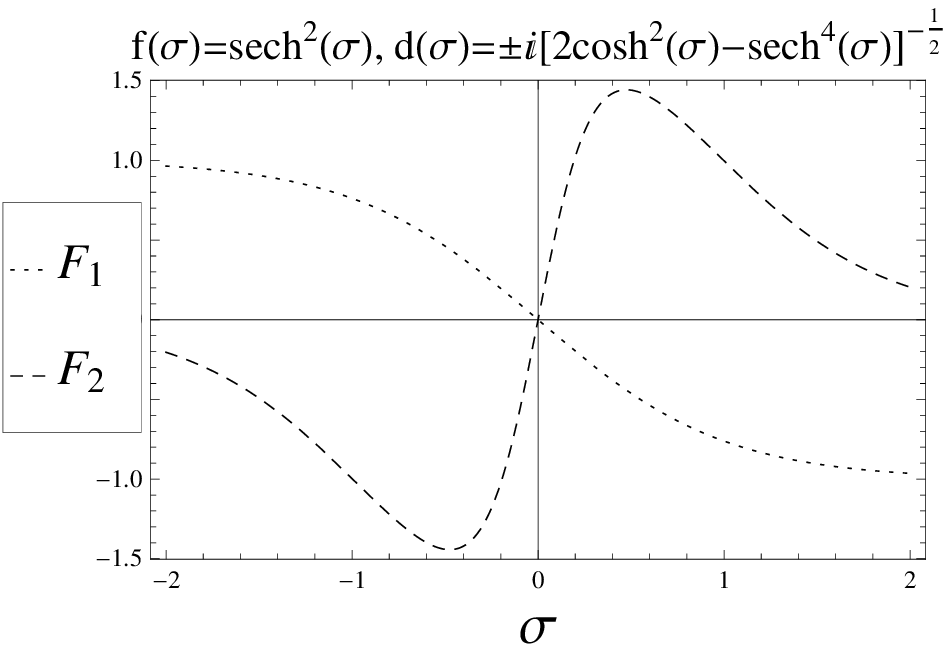} 
\caption{Comparing the coefficients appearing in the geodesic equation where $F_1 = f_{,\s}/2f$ and $F_2 = -ff_{,\s}/2F$.}
\label{fig:case2_F}
\end{center}
\end{figure}

The geodesics in presence of the decaying warp factor are shown in Fig. \ref{fig:case2_2}.
As mentioned earlier the non-symmetric components drastically change the energy density profile compared to the symmetric scenario. So it is expected that the geodesics to be modified.
Fig. \ref{fig:case2_2} shows that, for $d(\s)=0$, the geodesics run away from $\s=0$ and become finite  with increasing $\eta$ (similar results were found in \cite{SG-SK-HN}). Whereas in presence of the non-symmetric components, geodesics become confined within finite distance from $\s=0$. This behaviour can be understood (in the same line of reasoning as before), by comparing the effects of the coefficients appearing in the geodesic equation, from Fig. \ref{fig:case2_F}. Thus in our construction of thick branes in a non-symmetric bulk, time-like geodesics are confined for {\it both} growing and decaying warp factors.
As expected in the case of decaying warp factor, with imaginary $d(\s)$, $G_{{}_{AB}}u^Au^B$ is complex.

\section{Dynamic bulk spacetime}

In the previous section, we showed that the non-symmetric components can effectively confine the test particles at the location of the brane.
In this section, we study the cosmological scenarios in the context of 5 dimensional non-symmetric models. We assume here that, $\s=0$ is the location of the brane and the 5-dimensional spacetime is given by
\begin{center}
\be
g_{{}_{AB}}  = \le( \begin{array}{lllll} 
-a(\eta)  & 0  & 0 & 0 & b(\eta) \\
0 & a(\eta) & 0 & 0 & b(\eta) \\
0  & 0 & a(\eta) & 0 & b(\eta)  \\
0 & 0 & 0 & a(\eta)& b(\eta) \\
-b(\eta) &-b(\eta)& -b(\eta) & -b(\eta) & r^2    \\
\end{array} \ri) \la{eq:metric2}
\ee
\end{center}
where the 4D part contains the usual cosmological scale factor $a(\eta)$ and $b(\eta)$ is the corresponding non-symmetric component along the extra dimension. Here $\eta$ is the conformal time.
The corresponding line element is given by
\be
ds^2 = a(\eta) [-d\eta^2 + d\vec x^2] + r^2 d\s^2. \la{eq:lineelement2}
\ee
Note that, in this model, there is no notion of location of the localised energy density or brane.

Solving Eq. ({\ref{eq:2}}) for the above metric (\ref{eq:metric2}), the components of $\Gamma^C_{{}_{AB}}$ as function of $a(\eta)$ and $b(\eta)$ can be computed (see Appendix). Using these, Eq. (\ref{eq:constraint2}) leads to
\be
3\f{b^2}{a^2}a_{,\eta} + r^2\le(\f{a_{,\eta}}{a} + \f{b_{,\eta}}{b}\ri) = 0 . \la{eq:constraint4}
\ee
One can easily notice the similarity between Eq. (\ref{eq:constraint4}) and Eq. (\ref{eq:constraint3}). Here again we note that the results do not depend on the sign of $b(\eta)$.

The geodesic equations, obeying the timelike constraint, are given by
\br
\ddot{\eta} + \f{\dot a}{2a} (\dot\eta^2 + \sum_i\dot x_i^2)+ \f{r^2a^2bb_{,\eta}+2b^4a_{,\eta}-2ab^3b_{,\eta}}{r^4a^3-4ab^4}  (\dot\eta + \sum_i\dot x_i)^2 &&\nn \\ 
+ \, r^2\f{r^2a^2bb_{,\eta}-2b^4a_{,\eta}-2r^2ab^2a_{,\eta}-2ab^3b_{,\eta}}{r^4a^4-4a^2b^4} \dot\s^2 &=& 0, \la{eq:geo4} \\
\ddot x_i + \f{\dot a}{2a} \dot\eta\dot x_i + \f{r^2a^2bb_{,\eta}+2b^4a_{,\eta}-2ab^3b_{,\eta}}{r^4a^3-4ab^4}  (\dot\eta^2 + \dot\eta\sum_i\dot x_i)  &=& 0, \la{eq:geo5} \\
\ddot\s - \f{r^2a^2b_{,\eta}+2b^3a_{,\eta}-2ab^2b_{,\eta}}{2r^4a^3-8ab^4} \le[a\dot\eta^2 + 2b~\dot\eta(\sum_i\dot x_i+\dot\s)\ri]  &=& 0. \la{eq:geo6}
\er
Eqs (\ref{eq:geo4})-(\ref{eq:geo6}) reduces to standard FLRW scenario when $b(\eta)=0$.
We study the energy density and the geodesics for the two cases-- A) the de Sitter brane: i.e. $a(\eta)\propto 1/\eta^2$ where $\eta\in[-1,0]$ and B) power law brane i.e. $a(\eta) \propto \eta^m$ where $\eta\in[1,\infty)$.

\subsection{de Sitter brane}

\begin{figure}[!htb]
\begin{center}
\includegraphics[scale=.58]{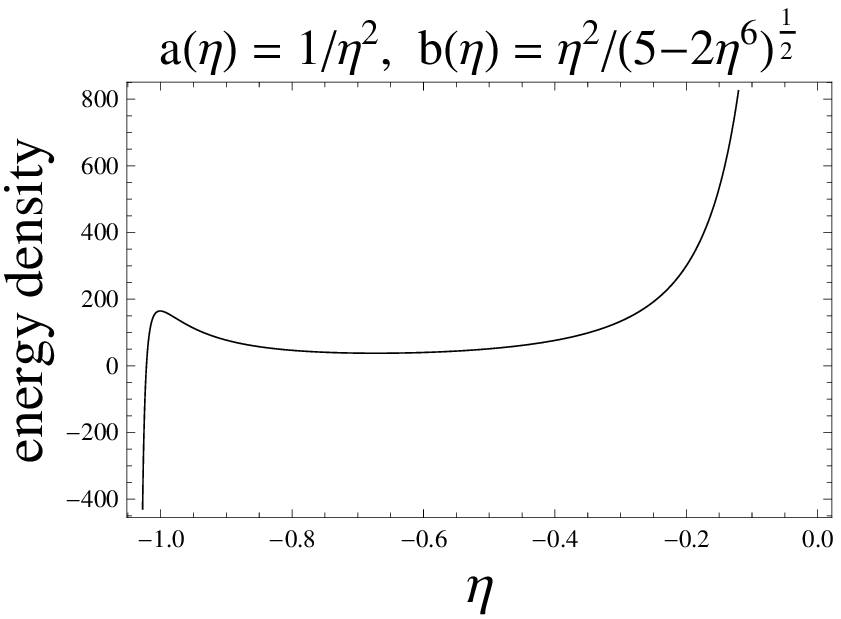} \hs{.5cm}
\includegraphics[scale=.58]{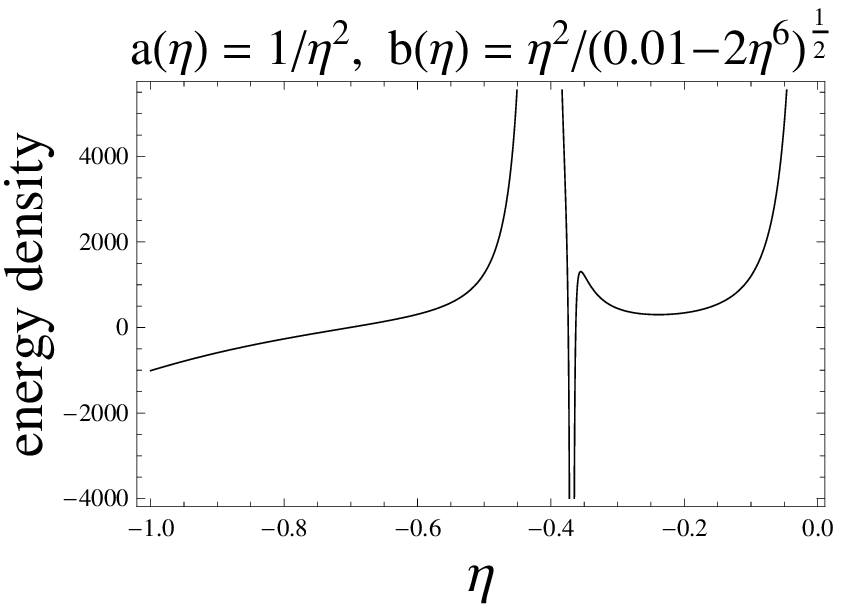} \hs{.5cm}
\includegraphics[scale=.58]{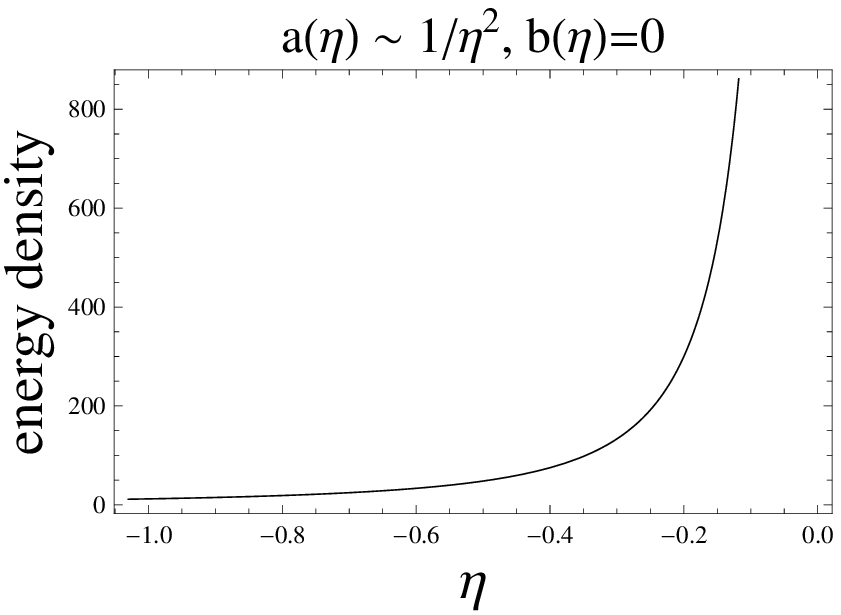} 
\caption{Plots of energy densities vs $\eta$ for de Sitter brane with $b(\eta)$ real ($C_3=5$), singular ($C_3=0.01$) at a particular $\eta$ and zero respectively.}
\label{fig:caset_1_G00}
\end{center}
\end{figure}

Substituting for $a(\eta)=1/\eta^2$ (where the proportionality constant is set to unity) in Eq. (\ref{eq:constraint4}), we get,
\be
b(\eta) = \f{r~\eta^2}{\sq{C_3r^2-2\eta^6}}
\ee
where $C_3$ is an integration constant. If $C_3>2/r^2$, $b(\eta)$ is real for all $\eta$.
However, for $C_3<2/r^2$, $b(\eta)$ is singular at $\eta = (C_3r^2/2)^{\f{1}{6}}$. In Fig.~\ref{fig:caset_1_G00}, we compare the energy densities for cases $C_3=5$ and $C_3=0.01$ with symmetric spacetime (i.e. $b(\eta)=0$). The non-symmetric property introduces an initial peak in the energy density which is absent in the symmetric case. However the asymptotic behaviours are similar in all the situations. The expression of energy density contains $(r^4a^2-4b^4)$ in the denominator which also appear in the Eqs (\ref{eq:geo4}-\ref{eq:geo6}). This leads to another singularity in the energy density apart from when $b(\eta)$ itself diverges.

\begin{figure}[!htb]
\begin{center}
\includegraphics[scale=.4]{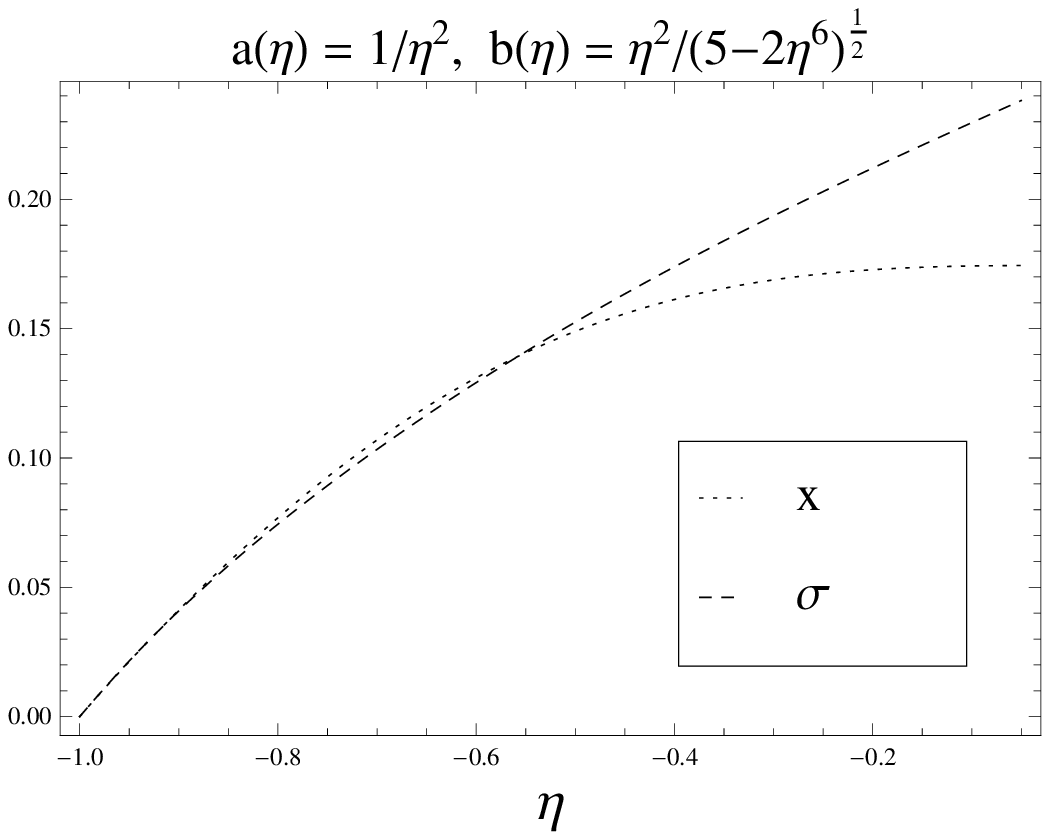} 
\includegraphics[scale=.55]{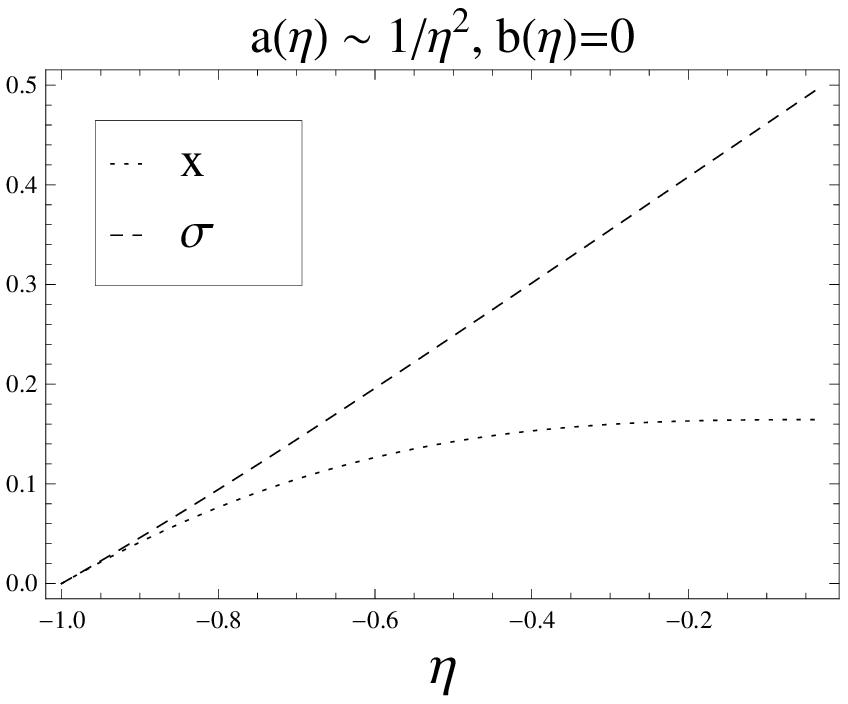} 
\caption{Plot of geodesics $x_i(\lambda)$ and $\sigma(\lambda)$ vs $\eta(\lambda)$ for de Sitter brane with $b(\eta)$ real and zero respectively. Initial conditions are $x_i(0)=\s(0)=0, \eta(0)=-1$ and $\dot x_i(0)=\dot\s(0)=1, \dot\eta(0)=\sq{5}$.}
\label{fig:caset_1_geo}
\end{center}
\end{figure}
The corresponding geodesics, shown in Fig. \ref{fig:caset_1_geo} with specific initial conditions, also behave similarly for the symmetric and non-symmetric models. Motion along the extra dimension slows down as $\eta$ increases whereas the components along $x_i$ increases monotonically. Note again that the geodesics are insensitive to initial conditions.

\subsection{Radiation dominated brane}

\begin{figure}[!htb]
\begin{center}
\includegraphics[scale=.58]{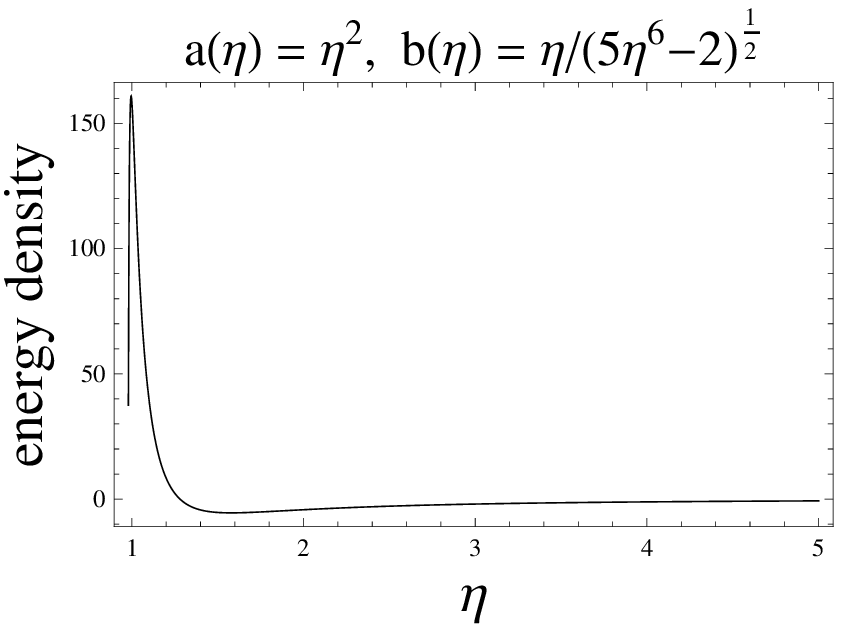} \hs{.5cm}
\includegraphics[scale=.58]{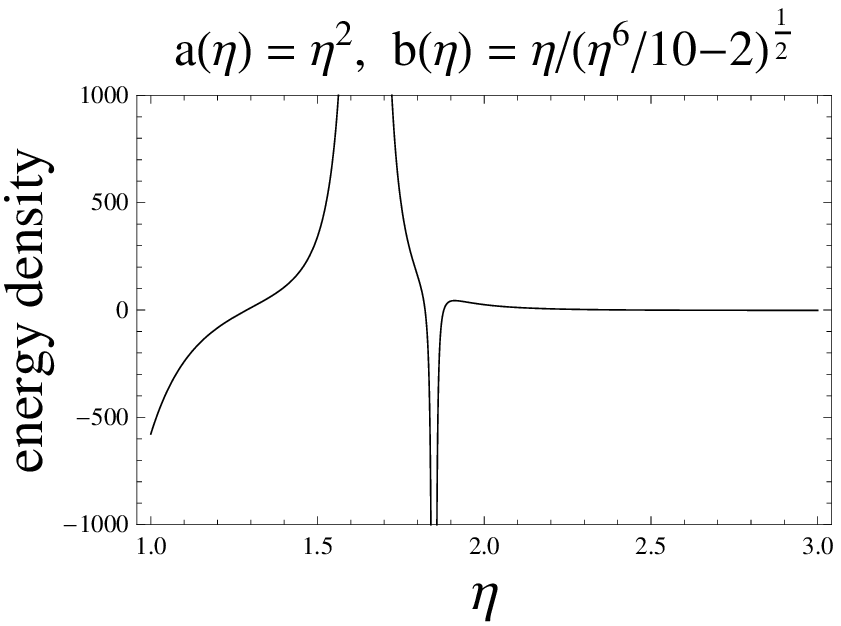} \hs{.5cm}
\includegraphics[scale=.58]{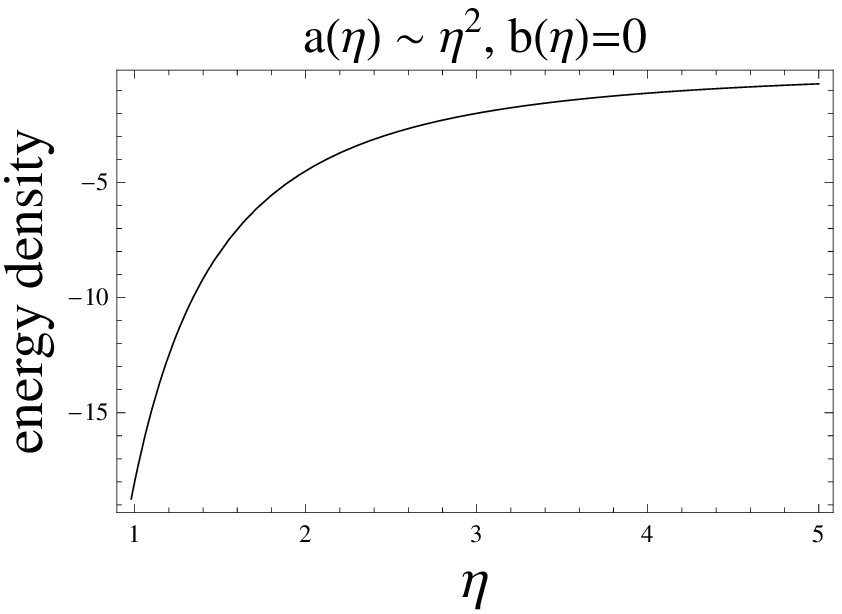} 
\caption{Plots of energy densities vs $\eta$ for Radiation dominated brane with $b(\eta)$ real, singular at a particular $\eta$ and zero respectively.}
\label{fig:caset_2_G00}
\end{center}
\end{figure}

\begin{figure}[!htb]
\begin{center}
\includegraphics[scale=.5]{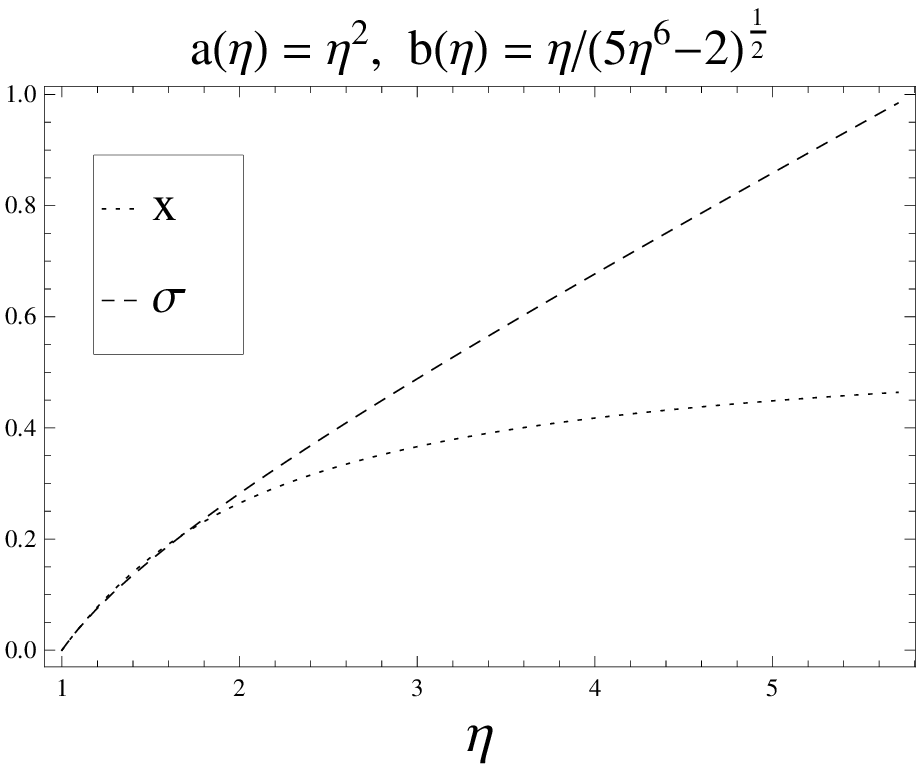} 
\includegraphics[scale=.48]{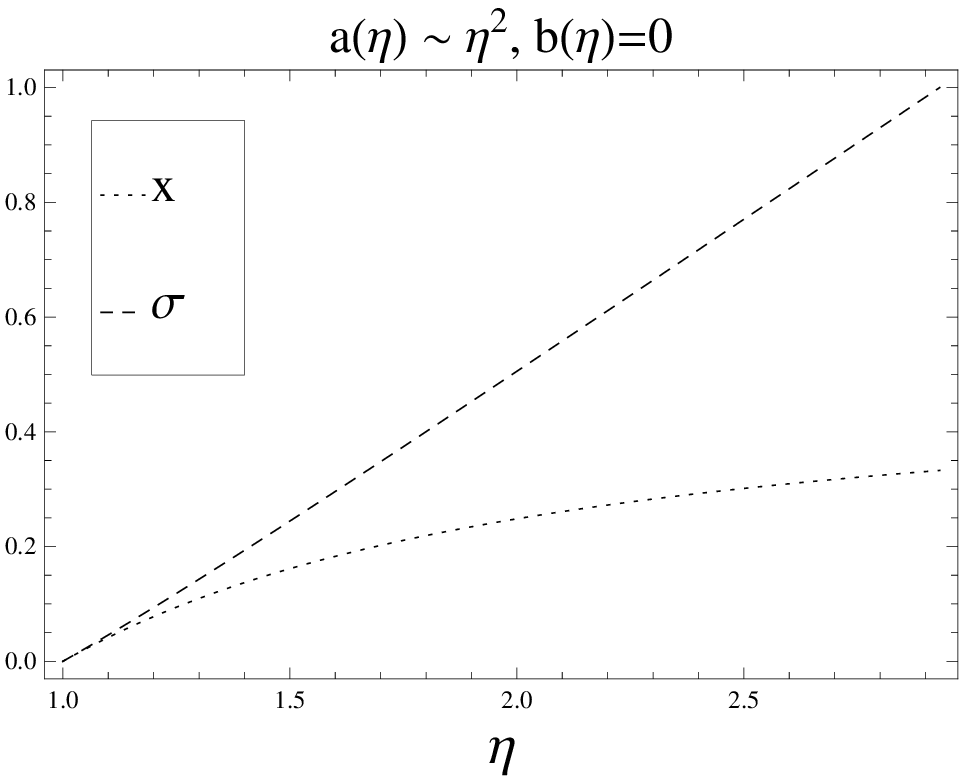} 
\caption{Plot of geodesics $x_i(\lambda)$ and $\sigma(\lambda)$ vs $\eta(\lambda)$ for Radiation dominated brane with $b(\eta)$ real and zero respectively.Initial conditions are $x_i(0)=\s(0)=0, \eta(0)=1$ and $\dot x_i(0)=\dot\s(0)=1, \dot\eta(0)=\sq{5}$.}
\label{fig:caset_2_geo}
\end{center}
\end{figure}

Substituting for $a(\eta) = \eta^m$ (here again, the proportionality constant is set to unity) in Eq. (\ref{eq:constraint4}) leads to
\be
b(\eta) = \f{r~\eta^{m/2}}{\sq{C_4r^2\eta^{3m}-2}}
\ee
where $C_4$ is the integration constant. For simplicity we choose $m=2$, that corresponds to a radiation dominated brane. Figs.~\ref{fig:caset_2_G00} and \ref{fig:caset_2_geo} shows the variations of energy density and geodesics for different values of $C_4$ and comparison w.r.t. $b(\eta)=0$ case. At early universe, a sharp peak in the energy density makes the cosmological evolution strikingly different from the symmetric case. However the geodesics are not significantly different in those two cases. Features shown are qualitatively similar to the previous model of a de Sitter brane because they both represent expanding 4D universes. The apparent distinction, in the asymptotic limit, between Figs.~\ref{fig:caset_1_G00} and \ref{fig:caset_2_G00} arises because de Sitter expansion results in a {\it big rip} whereas the radiation dominated universe becomes {\it flat} at large $\eta$. 

Note that one can further analyse these solutions for different values of $m$ which correspond to different cosmological scenarios.

\section{Search for warped dynamic bulk spacetime}

Let us investigate the possibility of constructing thick warped cosmological 3-branes embedded in 5D non-symmetric spacetime within the framework we have constructed here. We choose the following ansatz:
\begin{center}
\be
g_{{}_{AB}}  = \le( \begin{array}{lllll} 
-a(\eta)f(\s)  & 0  & 0 & 0 & b(\eta)d(\s) \\
0 & a(\eta)f(\s) & 0 & 0 & b(\eta)d(\s) \\
0  & 0 & a(\eta)f(\s) & 0 & b(\eta)d(\s)  \\
0 & 0 & 0 & a(\eta)f(\s) & b(\eta)d(\s) \\
-b(\eta)d(\s) &-b(\eta)d(\s)& -b(\eta)d(\s) & -b(\eta)d(\s) & r^2    \\
\end{array} \ri) \la{eq:metric3}
\ee
\end{center}
with the line element
\be
ds^2 = a(\eta)f(\s) [-d\eta^2 + d\vec x^2] + r^2 d\s^2. \la{eq:lineelement3}
\ee
Then Eqs. (\ref{eq:1}) leads to the following coupled equations:
\br
3\f{b^2}{a^2}a_{,\eta} + r^2\le(\f{f}{d^2}\ri)\le(\f{a_{,\eta}}{a} + \f{b_{,\eta}}{b}\ri) &=& 0, \la{eq:constraint5}\\
3\f{d^2}{f^2}f_{,\s} + r^2\le(\f{a}{b^2}\ri)\le(\f{f_{,\s}}{f} + \f{d_{,\s}}{d}\ri) &=& 0. \la{eq:constraint6}
\er
Only possible scenario where Eqs (\ref{eq:constraint5}) and (\ref{eq:constraint6}) can be solved is when 
\be
a(\eta) \propto b^2(\eta) \hs{1cm}\mbox{and}\hs{1cm} f(\s)\propto d^2(\s). \la{eq:2constraints}
\ee
Substituting Eq. (\ref{eq:2constraints}) in Eqs (\ref{eq:constraint5}) and (\ref{eq:constraint6})  leads to:
\be
a(\eta), b(\eta), f(\s), d(\s) = \rm{constants}, \la{eq:flatbulk}
\ee
i.e. a flat bulk spacetime. Thus thick brane solutions in a dynamic bulk spacetime is difficult to find in this construction. 
One interesting possibility would be to assume different functions as non-symmetric components. We hope to report on these aspects in a later article.

\section{Discussion}

Non-symmetric gravity models have been proposed as a natural modification or extension of GR.
In this work we have analysed, in detail, a yet unexplored aspect of non-symmetric geometry where warped thick branes are embedded in 5D spacetime. Here we summarise the key results.

\begin{enumerate}
\item We first introduced the field equations corresponding to non-symmetric geometry. We have considered models for which the anti-symmetric Ricci tensor (\ref{eq:2ndricci}) vanishes. This assumption considerably reduces the mathematical complexity in solving the field equations. 
Though relaxing that constraint may lead to richer class of geodesics which shall capture the essence of a non-vanishing 2nd Ricci tensor. 
To understand the physical implications of these models we have analysed the corresponding energy densities and geodesics. 
\item These models provide a natural mechanism for geodesic confinement. They also satisfy the 
Weak Energy Condition when the non-symmetric metric components are real. 
\item We have constructed stationary 5D bulk spacetime where thick branes with growing or 
decaying warp factors are embedded. The non-symmetric components are derived from Eqs. (\ref{eq:constraint2}) w.r.t. chosen warp factors. The timelike geodesics are analysed in both the cases. It is found that the geodesic component along the extra dimension is confined, by the geometry itself, near the location of the brane in both the scenarios with growing and decaying warp factors. This result is different from the corresponding symmetric spacetime cases where such confinement can only be achieved for growing warp factors. 
\item In the dynamical scenario, non-symmetric components are derived assuming standard FLRW 
brane embedded in a 5D spacetime. We considered de Sitter and radiation dominated expansion in the 4D spacetime. 
The corresponding energy densities show considerable departure from the symmetric models at early times. This suggests that the early universe dynamics is different. It may be interesting to look at the scalar power spectrum for these models \cite{ekpyrosis}. However the geodesic profiles are not drastically modified by the effect of non-symmetricity.
%
\item It is natural to ask whether there is an acceptable matter model that lead to thick-brane 5D non-symmetric space-times \cite{thick-brane,SG-SK-sols,SP-SK}. Numerically, we have shown the bulk energy density profiles, however, it is harder to obtain the same analytically for canonical scalar fields. This work is currently in progress. 
\end{enumerate}

To understand the role of non-symmetric properties of the spacetime on the curvature we need to study geodesic deviations. 
Further the geodesic and Raychaudhuri equations need to be solved simultaneously to investigate the kinematical quantities: expansion, rotation, and shear \cite{SG-SK-AD}. Such models could be constrained using the observational data \cite{Will}. 
We hope to report on these aspects elsewhere.

\section*{Acknowledgements}
This work is supported under the DST--Max Planck Partner Group on Gravitation and Cosmology and Ramanujan fellowship of DST, India. We thank S. Kar for useful discussions.

\begin{center}
\section*{Appendix}
\end{center}

The affine connections derived from Eq. (\ref{eq:2}) for stationary and dynamic models are given below.

\subsubsection{Stationary model}

\br
&& \Gamma^{\nu}_{\mu \nu} = -\Gamma^{\nu}_{\nu\mu} = \f{df_{,\s}}{2r^2f}, \,\, \nu\neq\mu, \nn \\
&&\Gamma^0_{04} = \Gamma^0_{40} = \f{r^2ff_{,\s} + 3d^2f_{,\s} - 2fdd_{,\s}}{2f(r^2f + 2d^2)}, \nn \\
&& \Gamma^0_{i4} = \Gamma^0_{4i} = \Gamma^4_{\nu\mu} = \Gamma^4_{\mu\nu} = -\Gamma^i_{4\mu} = -\Gamma^i_{\mu 4} = \f{d(df_{,\s} - 2fd_{,\s})}{2f(r^2f + 2d^2)}, \,\, \nu\neq\mu, \nn
\er
\br
&& \Gamma^i_{4i} = \Gamma^i_{i4} = \f{r^2ff_{,\s} + 3d^2f_{,\s} + 2fdd_{,\s}}{2f(r^2f + 2d^2)}, \Gamma^4_{00} = \f{f_{,\s}}{2r^2} + \f{d(df_{,\s} - 2fd_{,\s})}{r^2(r^2f + 2d^2)}, \nn \\
&& \Gamma^4_{ii} = - \f{f(4dd_{,\s} + r^2f_{,\s})}{2r^2(r^2f + 2d^2)}.  \nn \\
\er

\subsubsection{Dynamical model} 

\br
&& \Gamma^0_{00} = \f{a_{,\eta}}{2a}, \,\,\Gamma^0_{ij} = \Gamma^0_{ji} = \Gamma^0_{i0} = \Gamma^0_{0i} =\f{r^2a^2bb_{,\eta} + 2b^4a_{,\eta} - 2ab^3b_{,\eta}}{a(r^4a^2-4b^4)} = \f{b}{a}\Gamma^0_{ii},\, i\neq j, \nn \\
&& \Gamma^0_{04} = - \Gamma^0_{40} = \f{r^2a^2b(r^2a_{,\eta}+7bb_{,\eta}) - 2r^4a^3b_{,\eta} - 2ab^3(r^2a_{,\eta}+3bb_{,\eta}) +  6b^5a_{,\eta}}{2a^2(r^4a^2 - 4b^4)}, \nn \\
&& \Gamma^0_{i4} = - \Gamma^0_{4i} = \f{r^2a^2b(r^2a_{,\eta}+5bb_{,\eta}) - r^4a^3b_{,\eta} - 2ab^3(r^2a_{,\eta}+3bb_{,\eta}) +  2b^5a_{,\eta}}{2a^2(r^4a^2 - 4b^4)}, \nn 
\er
\br
&& \Gamma^0_{44} = \f{r^2b(r^2a^2b_{,\eta} - 2b^3a_{,\eta} -2ab(r^2a_{,\eta}+bb_{,\eta}))}{a^2(r^4a^2 - 4b^4)} ,\nn \\
&& \Gamma^i_{00} = \f{b(r^2a^2b_{,\eta} + 2b^3a_{,\eta} - 2ab^2b_{,\eta})}{a(r^4a^2 - 4b^4)}, \nn \\
&& \Gamma^i_{0i} = \Gamma^i_{i0} = \f{r^2a^2(r^2a_{,\eta}+bb_{,\eta}) - 2b^3(ba_{,\eta} + ab_{,\eta})}{2a(r^4a^2 - 4b^4)}, \nn \\
&& \Gamma^i_{0j} = \Gamma^i_{j0} = \f{b(r^2a^2b_{,\eta} + 2b^3(ba_{,\eta}-ab_{,\eta}))}{2a(r^4a^2 - 4b^4)}, \nn 
\er
\br
&& \Gamma^i_{4i} = -\Gamma^i_{i4} = \f{b(r^2a^2(r^2a_{,\eta}+bb_{,\eta}) - 2b^3(ba_{,\eta} + ab_{,\eta}))}{2a^2(r^4a^2 - 4b^4)}, \nn \\
&& \Gamma^i_{4j} = -\Gamma^i_{j4} = \f{b^2(r^2a^2b_{,\eta} + 2b^3(ba_{,\eta}-ab_{,\eta}))}{2a^2(r^4a^2 - 4b^4)},\, i\neq j, \nn \\
&& \Gamma^i_{04} = -\Gamma^i_{40} = \f{r^4a^3b_{,\eta} - r^2a^2b(r^2a_{,\eta}+3bb_{,\eta}) - 2b^4(ba_{,\eta} - ab_{,\eta})}{2a^2(r^4a^2 - 4b^4)}, \nn 
\er
\br
&& \Gamma^4_{i0} = -\Gamma^4_{0i} = \f{r^2a^2b_{,\eta} + 2b^2(ba_{,\eta}-ab_{,\eta})}{2a^2(r^4a^2 - 4b^4)}, \nn \\
&& \Gamma^4_{40} = \Gamma^4_{04} = \f{b(r^2a^2b_{,\eta} - 2b^3a_{,\eta} - 2ab(r^2a_{,\eta}+bb_{,\eta}))}{2a(r^4a^2 - 4b^4)}, \nn \\
&& \Gamma^4_{4i} = \Gamma^4_{i4} = - \f{b(r^2a^2b_{,\eta} - 2b^2(r^2a_{,\eta}+bb_{,\eta}))}{2a(r^4a^2 - 4b^4)}. \nn \\
\er



\end{document}